\documentclass[twocolumn]{aastex62}

\newcommand{\Ic}{I_{\rm c}}
\usepackage{color}
\renewcommand{\ion}[2]{{#1}\,{\sc #2}}

\def\arcsec{\hbox{$^{\prime\prime}$}}
\def\arcsec{\ensuremath{^{\prime\prime}}}

\received{10 April 2018}
\revised{}
\accepted{8 May 2018}

\submitjournal{ApJ}

\shorttitle{Understanding the HMI pseudocontinuum in white-light solar flares}
\shortauthors{\v{S}vanda et al.}

\begin{document}

\title{Understanding the HMI pseudocontinuum in white-light solar flares}

\correspondingauthor{Michal \v{S}vanda}
\email{michal@astronomie.cz}

\author[0000-0002-6345-1007]{Michal \v{S}vanda}
\affiliation{Astronomical Institute of the Czech Academy of Sciences, 
Fri\v{c}ova  298, 25165 Ond\v{r}ejov, Czech Republic}
\affiliation{Astronomical Institute, Charles University, V~Hole\v{s}ovi\v{c}k\'ach 2, 18000 Praha, Czech Republic}

\author{Jan Jur\v{c}\'ak}
\affiliation{Astronomical Institute of the Czech Academy of Sciences, Fri\v{c}ova  298, 25165 Ond\v{r}ejov, Czech Republic}

\author{Jana Ka\v{s}parov\'a}
\affiliation{Astronomical Institute of the Czech Academy of Sciences, Fri\v{c}ova  298, 25165 Ond\v{r}ejov, Czech Republic}

\author[0000-0002-7791-3241]{Lucia Kleint}
\affiliation{University of Applied Sciences and Arts Northwestern Switzerland, Bahnhofstrasse 6, 5210 Windisch, Switzerland}
\affiliation{Kiepenheuer-Institut f\"{u}r Sonnenphysik, Sch\"{o}neckstr. 6, 79104 Freiburg, Germany}

\begin{abstract}
We analyse observations of the X9.3 solar flare (SOL2017-09-06T11:53) observed by SDO/HMI and Hinode/SOT. Our aim is to learn about the nature of the HMI pseudocontinuum $\Ic$ used as a proxy for the white-light continuum. From model atmospheres retrieved by an inversion code applied to the Stokes profiles observed by the Hinode satellite we synthesise profiles of the \ion{Fe}{i} 617.3~nm line and compare them to HMI observations. Based on a pixel-by-pixel comparison we show that the value of $\Ic$ represents the continuum level well in quiet-Sun regions only. In magnetised regions it suffers from a simplistic algorithm that is applied to a complex line shape. During this flare both instruments also registered emission profiles in the flare ribbons. Such emission profiles are poorly represented by the six spectral points of HMI, the used algorithm does not account for emission profiles in general, and thus the derived pseudocontinuum intensity does not approximate the continuum value properly. 

\end{abstract}


\section{Introduction}

The Sun belongs to the group of active stars where various kinds of active phenomena are observed, such as solar flares. The continuous observations by several satellites, including images taken every few seconds by SDO, allow us to investigate flares in high resolution and over long periods.

In the current paradigm, flares on the Sun and Sun-like stars are a consequence of a very fast change in the magnetic field topology. The topology change -- the magnetic reconnection -- usually occurs  in the coronal loops above magnetic regions \citep[e.g.][]{2011LRSP....8....6S}. The phenomenon of a `flare' consists of formation of particle beams, intense local atmospheric heating, and release of coronal mass ejections. Solar flares are sources of electromagnetic radiation at all wavelengths from radio waves to gamma-rays. Although they are prominent in EUV and X-ray emission, the majority of the radiated flare energy emerges at visible and UV wavelengths
\citep{2006JGRA..11110S14W,2011A&A...530A..84K} and originates in the lower atmosphere, namely in the form of continuum
radiation \citep[e.g.][]{1989SoPh..121..261N}.

Despite the fact that most of our information about solar flares has been devised from analyses of observations obtained either in chromospheric emission lines, EUV, X-ray, and radio emission,  
the first ever observed solar flare was seen by naked eye in the white light \citep{1859MNRAS..20...13C,1859MNRAS..20...15H}. 
White-light flares -- flares with emission in the visible continuum --
\citep[e.g.][]{1966SSRv....5..388S,1989SoPh..121..261N}
are probably not a rare phenomenon. It is now known that the visible continuum enhancement appears during flares as weak
as GOES C-class 
\citep{2003A&A...409.1107M,2008ApJ...688L.119J}. Sometimes the enhancement relative to 
the photospheric brightness can be low, typically few tens of percent, and it can be challenging to detect even in X-class flares \citep{2014ApJ...783...98K}.
\begin{figure}
\plotone{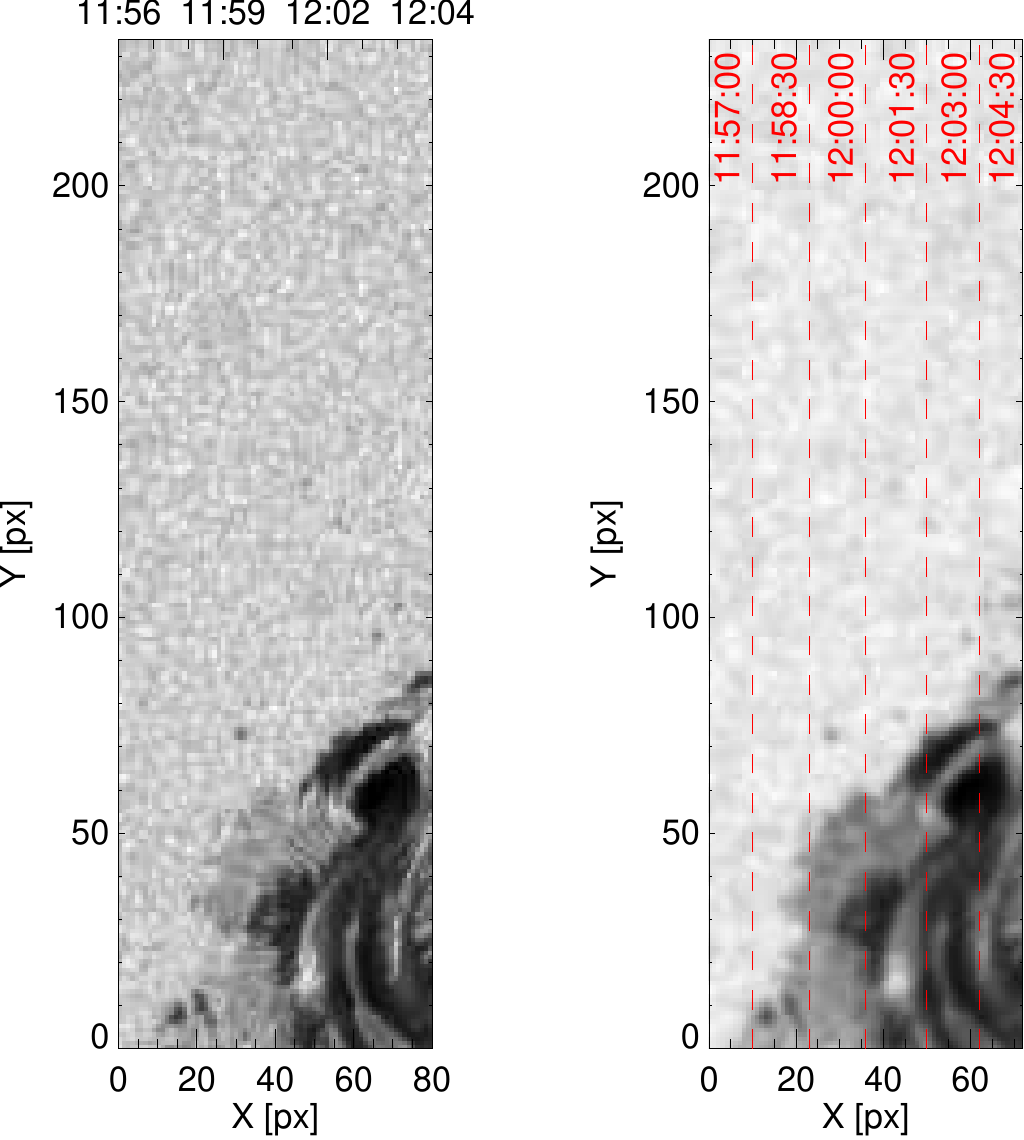}
\caption{The co-aligned fields of view. Left: Hinode/SOT SP scan displayed at continuum spectral point of intensity profile. The scanning was performed in the horizontal direction 
indicated by an additional time axis on the top. Right: the co-aligned pseudoscan from HMI. The red dashed lines indicate the transition between the HMI frames in different times (that are also given in red letters). Note that the outermost filter position was used in this plot. The times represent UT of September 6, 2017.}
\label{fig:FOV_compare}
\end{figure}

Usually, the increase in the visible-light continuum is interpreted as a combination of the 
Paschen continuum (due to hydrogen recombination) and photospheric, H$^{-}$, component. 
Recent observations
and dedicated models also show that the hydrogen Balmer  \citep{2016ApJ...816...88K,2017ApJ...836...12K} and
Paschen  \citep{2017ApJ...847...48H} continuum emission are indeed present in the white-light flares. 
Energy transport from emission in the Balmer continuum in higher atmospheric layers towards the photosphere could also lead to a further visible continuum enhancement through radiative heating of the lower atmosphere, i.e. to the increase of H$^{-}$ emission \citep{1989SoPh..124..303M}.

Visible-light emission was observed by many instruments in the past. In order to properly study this emission, long-term and high-resolution data are needed. Synoptic experiments such as the Helioseismic and Magnetic Imager \citep{2012SoPh..275..229S} on-board of the Solar Dynamics Observatory, HMI/SDO, then constitute a holy grail for such studies. 

HMI data products have been used in past studies to investigate white-light flares.
For instance, a statistical analysis of \citet{2016ApJ...816....6K} shows a correlation between hard X-ray
flux at 30~keV and white-light flux at 617.3~nm summed over hard X-ray flare ribbons.  Similarly, \cite{2017ApJ...851...91N} studied statistically 50 flares recorded in the HMI archive in order to compare the scaling relations with those derived for stellar flares, such as the energies of the flares. Additionally, recent studies suggest that hard X-ray and white-light emission come from the same volume
\citep{2012ApJ...753L..26M,2015ApJ...802...19K}.

In this paper, we present a case study of a X9.3-class flare that occurred on September 6, 2017 (SOL2017-09-06T11:53). We take the advantage of the observations taken by the Solar Optical Telescope \citep[SOT,][]{2008SoPh..249..167T} aboard the Hinode satellite \citep{2007SoPh..243....3K} as this event triggered the flare mode of this satellite. Namely, we use the spectropolarimetric (SP) data from Hinode/SOT. To our knowledge, it was for the first time that Hinode SP raster scan captured a white-light flare ribbon. The availability of both the HMI observations and Hinode spectral scans make this flare an ideal target to study the nature of white-light flare emission recorded in the HMI $\Ic$ pseudocontinuum. 

We compare synthetic and observed \ion{Fe}{i} 617.3~nm line profiles and discuss reliability of HMI pseudocontinuum values in different structures, i.e. umbra, penumbra, quiet Sun, and flare region.

\section{Observations and data processing}\label{obs} 

The studied event was the largest flare recorded in the 24th solar cycle and originated from the fast-growing active region NOAA~12673 that appeared as it rotated across the visible hemisphere of the Sun. The active region emerged on August 27 as a single spot surrounded by a plage region. A very rapid emergence of a new flux started on September 3 and a very complex magnetic topology in the region including a $\delta$-configuration was formed. The flux emergence rate was enormous, peaking at ${1.12}_{-0.05}^{+0.15}\times {10}^{21}\,\mathrm{Mx}\,{\mathrm{hr}}^{-1}$, which is the largest recorded value ever \citep{2017RNAAS...1...24S}. \cite{2017ApJ...849L..21Y} speculated that the strong flaring activity of the region that peaked with the X9.3 flare was due to the peculiar configuration of the magnetic field including a high shear.

HMI observed in a regular mode during the analysed flare. We use the $\Ic$ intensitygrams available from JSOC\footnote{jsoc.stanford.edu} archive at a cadence of 45~s. Their spatial sampling is 0.5\arcsec{} per pixel. In Sect.~\ref{HMI_Ic}, we summarise the method used for derivation of the $\Ic$ product. For the analysed flare, there are also available the Stokes $I$, $Q$, $U$, and $V$ filtergrams of the \ion{Fe}{i} 617.3~nm line taken with a cadence of 90~s with a spatial sampling of 0.5\arcsec{} per pixel.

The flaring region was captured by the raster scan of the Hinode spectropolarimeter. The scan took approximately nine minutes and captured a part of active region 12673, covering most of the area where the ribbons of the flare were detected. The Stokes profiles of the two \ion{Fe}{i} lines at 630.15~nm and 630.25~nm were observed. In this particular observational mode, the slit had a length equivalent to 123\arcsec{} and spatial sampling of 0.32\arcsec{} along the slit. The raster consists of 146 slit positions taken with a cadence of 3.8~s per slit position and with slit-step size (the spatial sampling) of 0.3\arcsec{}. The data were calibrated with the standard routines available in the Hinode SolarSoft package. The resulting noise level is $1.4 \times 10^{−3}~I^c_{QS}$ (where $I^c_{QS}$ stands for the intensity of the quiet Sun) for the Stokes $I$ profile and $1.8 \times 10^{−3}~I^c_{QS}$ for the Stokes $Q$, $U$, and $V$ profiles. 

\subsection{HMI data products}
\label{HMI_Ic}

HMI observes the Sun at a relatively high cadence (standard data products are available each 45~s) continuously since 2010. Among the standard data products, the derived pseudocontinuum intensity, $\Ic$, is seemingly the most suitable for studies of the visible-light emission in flares. Despite being called this name, $\Ic$ is \emph{not} a measured continuum intensity. 

HMI scans the \ion{Fe}{i} 617.3~nm photospheric absorption line and measures the intensity in different polarisation states. In a standard regime, the sequence of filtergrams is obtained by scanning the spectral line in six wavelength positions 7.6~pm apart with a variety of polarimetric states. The raw unprocessed filtergrams in polarised light are the level~0 series, which is only available upon request. After flat-fielding and application of other calibration procedures, primary filtergrams are stored as level~1 data, available from JSOC. From the level~1 data various data products are computed. Among others, it is the derived pseudocontinuum intensity $\Ic$. 

The derived pseudocontinuum intensity stored within the series {\tt hmi.Ic\_45s} at JSOC is computed using a fast ``MDI-like algorithm'' from a set of level 1 filtergrams assuming a certain shape of the spectral line as \citep[from][]{2012SoPh..278..217C}
\begin{equation}
\Ic \approx \frac16 \sum\limits_{j=0}^5 \left[ I_j + L_d \exp\left(-\frac{(\lambda_j-\lambda_0)^2}{\sigma^2} \right)  \right],
\label{eq:ic}
\end{equation}
where $I_j$ are the intensities in each of six filtergrams scanning the iron line profile, $\lambda_j$ is the corresponding wavelength, and $\sigma$ is the line width. \cite{2012SoPh..278..217C} point out that the line-width value is not part of the process to derive $\Ic$ because the simplistic algorithm determining the line width fails frequently. Instead, a ``standardised'' line width is taken, derived usually in the quiet-Sun region. Moreover, the varying line width e.g. in the magnetised regions may lead to a bias in determining the continuum intensity.

HMI data products go far beyond line-of-sight observables. A set of Stokes $I$, $Q$, $U$, $V$ six-position profiles is derived with a cadence of 720~s from level 1 polarised-light filtergrams and stored in {\tt hmi.S\_720s} series. For special events a shorter-cadence 
{\tt hmi.S\_135s} series can be derived and even a shorter averaging time (90~s) was introduced for an experimental series \cite[see][]{2017ApJ...839...67S}. The data from {\tt su\_yang.HMISeriesLev1pa90Q} series available from JSOC2\footnote{jsoc2.stanford.edu} together with the standard $\Ic$ series are used in this study. 

The quiet-Sun regions were used to determine the noise level in the polarisation profiles of HMI measurements. There we can expect polarisation signals close to zero. The analysis of histograms of $Q$, $U$, and $V$ signals confirmed this expectation; the values are normally distributed with a zero mean value and the $\sigma$ values of these distributions correspond to the resulting noise level. The value that was found is comparable to the Hinode/SP measurements and is around $1.8 \times 10^{−3}~I^c_{QS}$ for the $Q$, $U$, and $V$ filtergrams.

\subsection{Modelling of the HMI 617.3 nm line from Hinode observations}\label{modelling}

Both HMI and SOT  observe the solar photosphere, yet in different spectral lines. Despite the fact that the formation heights of all three considered lines are similar, a direct comparison is not possible, e.g. due to the different magnetic and temperature sensitivities of the \ion{Fe}{i} 617.3~nm line and the \ion{Fe}{i} 630.15~nm and 630.25~nm lines. However, given the similar formation height of the observed spectral lines, we may use the model atmosphere derived from the Hinode observations to synthesise the Stokes profiles of the HMI line.

Using the SIR code \citep{1992ApJ...398..375R}, we inverted the Hinode/SP observations in a pixel-by-pixel manner and obtained the model atmosphere covering optical depths $\log\tau=-3.8$ to $\log\tau=1$. In the inversion scheme, we assumed that the physical parameters are constant with height in the atmosphere apart for the temperature that was allowed to change in five optical depths and the temperature stratification in a finer grid was obtained by interpolation. We take into account the spectral PSF of the Hinode/SP instrument, do not allow for macro-turbulence in the model atmosphere, and do not take into account any stray light.

\begin{figure}
\includegraphics[width=3.5cm]{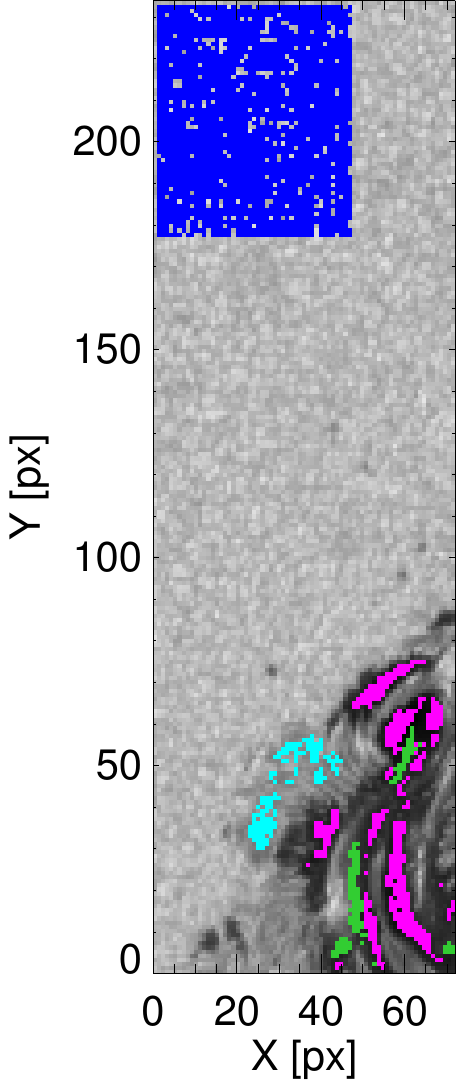}
\caption{The analysed field of view with various regions indicated. Dark blue denotes the regions of the quiet Sun, light blue the penumbra, pink the umbra, and green the regions where the emission profiles are present.}
\label{fig:FOV}
\end{figure}
To obtain the profiles of the HMI line, we used the resulting models of atmospheres in pixels, where the observed Hinode/SP lines were successfully reproduced (criterion based on the $\chi^2$ value of the SIR fit). These model atmospheres were used to synthesise the Stokes profiles of the \ion{Fe}{i} 617.3~nm line using again the SIR code. The resulting models and their relevance to realistic conditions in the solar atmosphere will be discussed in detail in Jur\v{c}\'ak et al. (in preparation).

We stress that the use of the synthesis of the \ion{Fe}{i} 617.3~nm line from the inverted atmosphere is absolutely necessary to properly understand the nature of HMI $\Ic$. As we will show later, the investigation of the HMI spectral scans without knowing the detailed line shape might lead to biased interpretations of HMI data products.

\subsection{Comparison of SOT and HMI observations}
\label{data_processing}

The data obtained by HMI and Hinode are of a very different kind. The HMI observations are filtergrams with a limited spectral resolution, whereas Hinode/SP observations are raster scans. Their spectral resolution is large, however the rasterised image does not contain simultaneous information all over the captured region. Furthermore, the raster takes $\sim 9$~min, i. e. much longer than the accumulation time of HMI series used here and also longer than the typical timescale of the evolution of the flare ribbons.

In order to compare those two sets of observations, we constructed a pseudoscan from the HMI dataset. First, the HMI filtergrams and $\Ic$ frames were co-aligned to the field of view of the Hinode/SP scan. In the case of HMI maps we used the outermost filtergram from the line centre, which in our case was always the filtergram in the red wing or $I_5$ from Eq.~(\ref{eq:ic}). We manually identified distinct features (such as the pores in the upper part of field of view or various contrast features in the sunspot regions), measured their pixel coordinates in both data products and determined the parameters of a linear transform to register the HMI filtergram to Hinode frame in the continuum. The transform included translation, rotation, scale change, and shear. Their values were determined by the least-squares fitting. By blinking the Hinode/SP scan and the remapped HMI data products we made sure that the co-alignment was correct within one pixel accuracy. 

Second, in the Hinode/SP scan the column coordinate indicates not only the position over the field of view in the horizontal direction but also the time when the spectrograph slit was positioned there. Therefore, for each column in Hinode/SP frame we knew the time of observation and searched for the HMI frame closest to that time from the co-aligned HMI series. Then we stacked the columns to form a raster similar to Hinode/SP scan. As a final step, the Hinode/SP spatial resolution was degraded to match that of the HMI data. Figure~\ref{fig:FOV_compare} shows the comparison of the Hinode/SP scan and the HMI pseudoscan.

The wavelength scales of the synthetic \ion{Fe}{i} 617.3~nm line profiles and the observed HMI line profiles do not match. There are many uncertainties in the position of the line centre of these profiles, e.g., due to the satellite motion. Instead of trying to understand all the effects leading to wavelength shifts, we rather fit the spectral profiles by using the least-squares method. In order to do that, the synthetic and observed \ion{Fe}{i} 617.3~nm line profiles were first normalised to the local continuum intensity, which was determined from the outermost point in the spectral coverage of the given instrument and averaged over the nearby quiet-Sun region.

To account for the HMI throughput, six HMI transmission profiles $F_j (\lambda)$ \citep[see e.g.][]{2014SoPh..289.3531C} were applied to the synthetic Stokes $I_\lambda(\lambda)$ profiles. Index $j$ indicates the corresponding filter of a given HMI spectral point. Then, we numerically minimised the function

\begin{equation}
\chi^2= \sum\limits_{j=1}^6 \left\{I_j(\lambda_j+\Delta\lambda) - A \int\limits_{-\infty}^{+\infty} \left[ I_\lambda (\lambda') F_j (\lambda'-\lambda_j) d\lambda' \right]  \right\}^2,
\label{eq:fit}
\end{equation}
where $I_j$ are pixel intensities from HMI filtergrams. The wavelength shift $\Delta\lambda$ and the multiplicative factor $A$ are free parameters to be determined by the minimisation. In this approach we assume that the two instruments may have different quantum efficiencies, $A$,  and their wavelength scales may be mutually shifted by $\Delta\lambda$. 

The pixel-to-pixel fitting will always result in the best match of the profiles in the least-squares sense. However, one needs to keep in mind that the observations from both instruments do not necessary represent the same place at the same time on the Sun. A slight spatial misalignment or temporal evolution of structures may lead to differences in the spectral-line profiles. To ``calibrate'' HMI to the synthetic 617.3~nm spectral line from Hinode we use the pixel-by-pixel fits restricted only to the quiet-Sun region and derive general trends in the fitting parameters. The proper value of the multiplicative factor $A$ from (\ref{eq:fit}) was determined as an average over the quiet-Sun region because we expect that this factor is due to the differences in instruments and should be constant. In our case $A=1.024$.

On the other hand, a wavelength shift $\Delta\lambda$ may evolve in time due to the motion of the satellite. A column-wise average of $\Delta\lambda$ oscillates about a trend in time with root-mean-square value of about 0.17~pm with a mean of about 5.65~pm. In the subsequent analysis we use only a smooth trend that is obtained by fitting a second order polynomial in time (that is over the columns in the field-of-view) to realisations of $\Delta\lambda$ again in the quiet-Sun region. The typical values of the parameters are then used in a direct comparison of all HMI and Hinode pixels. 

\begin{figure}
\plotone{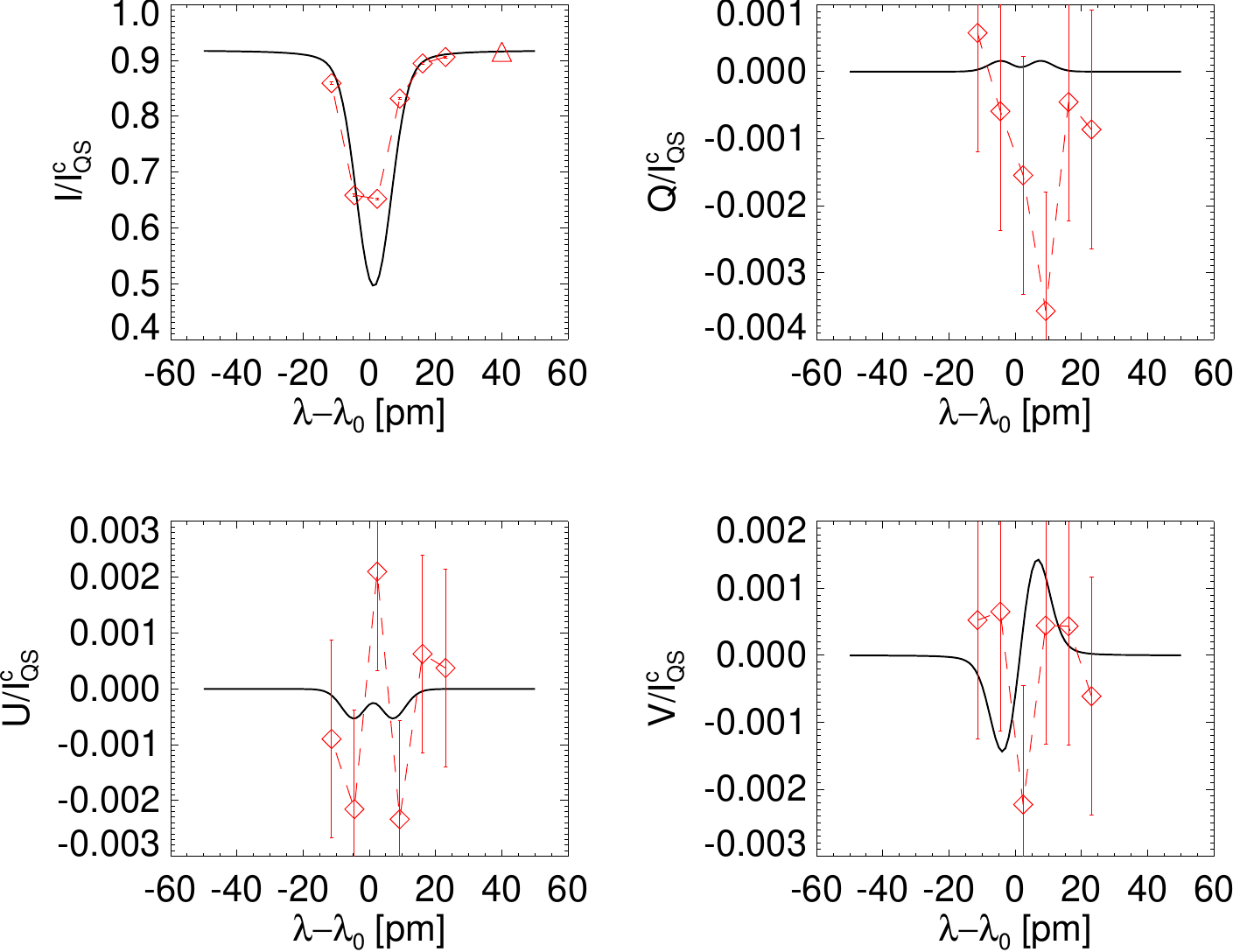}
\caption{An example of Stokes profiles for the quiet Sun. The panels indicate the $I$, $Q$, $U$, and $V$ profiles. The solid black line represent the %synthesis of the 
synthetised 617.3~nm spectral line based on the atmosphere derived from inversions of Hinode/SP data, smoothed with a Gaussian of FWHM=7.6~pm (to approximate the HMI transmission filters). The red diamonds indicate the spectral points derived from HMI data series and the vertical red lines mark the estimated noise level. The red triangle denotes the $\Ic$ HMI pseudocontinuum.}
\label{fig:profiles_quiet}
\end{figure}

\begin{figure}
\plotone{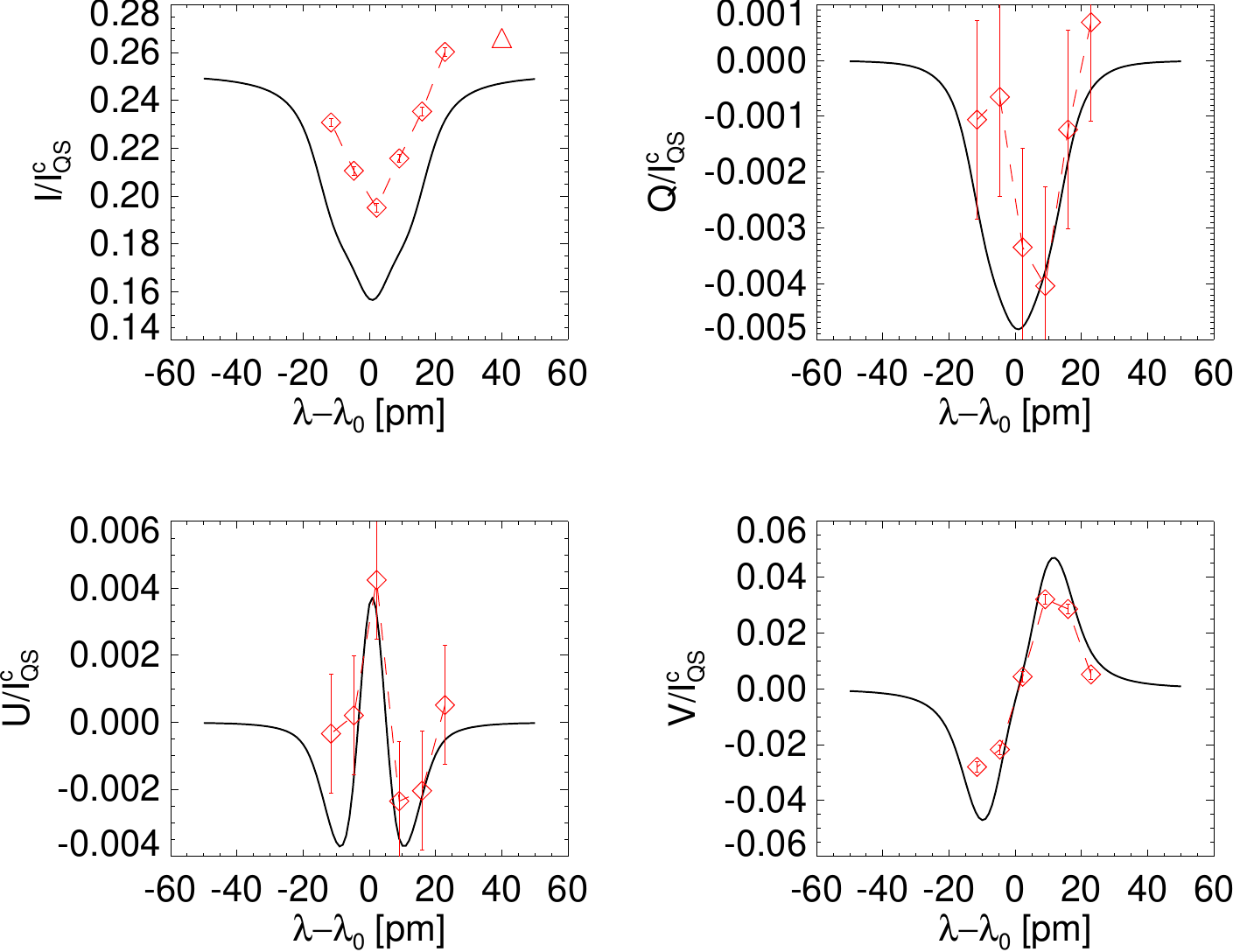}
\caption{The Stokes profiles for an example umbral point. Symbols are the same as in Fig.~\ref{fig:profiles_quiet}.}
\label{fig:profiles_umbra}
\end{figure}

\begin{figure}
\plottwo{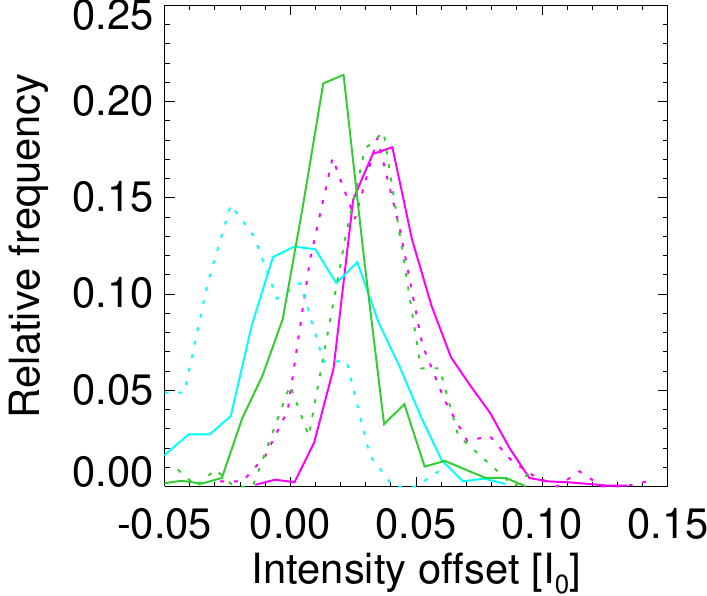}{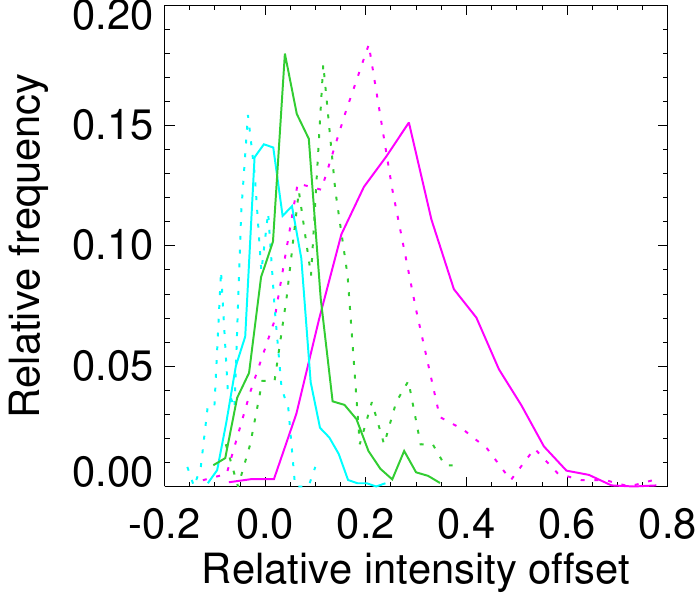}
\caption{Histograms of intensity offsets for profiles in the umbra (pink), penumbra (light blue), and emission regions (green). The solid lines correspond to the HMI filtergrams, the dashed lines to $\Ic$. Left: absolute offset values, right: relative offset values with respect to the model.}
\label{fig:histograms}
\end{figure}

\begin{figure*}
\plotone{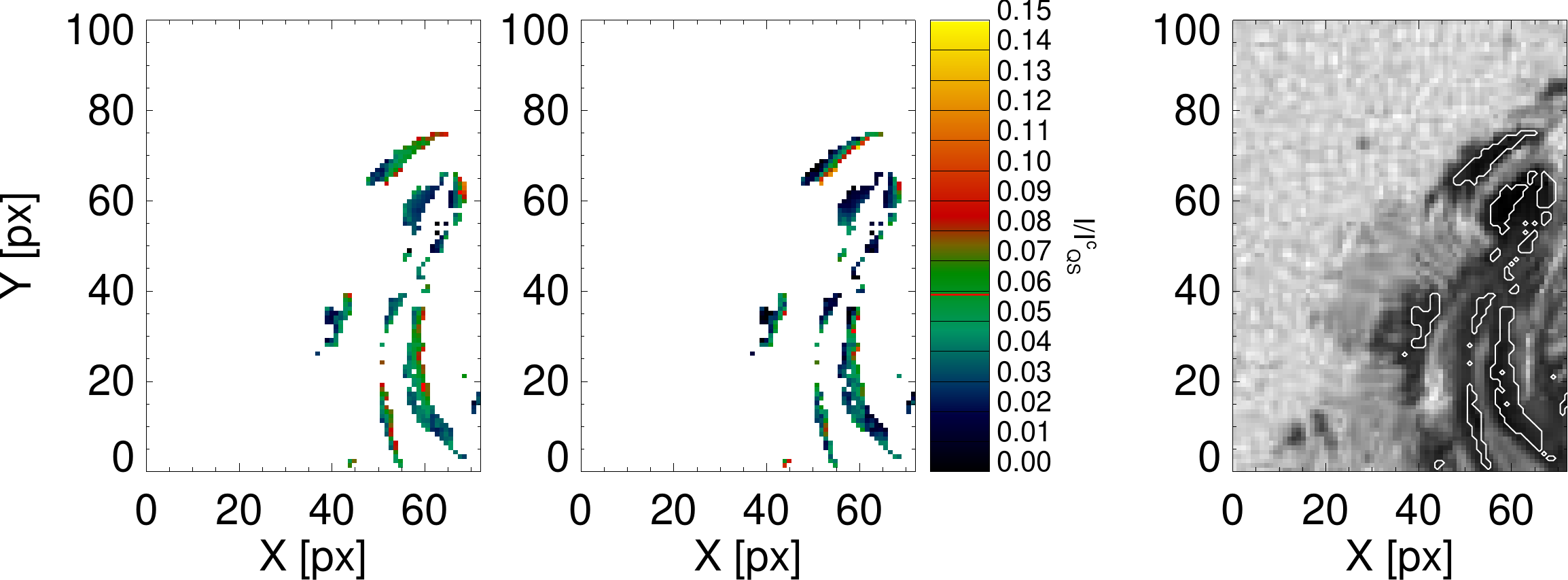}
\caption{The maps of intensity offsets in umbral regions in the spectral line (left) and $\Ic$ pseudocontinuum (middle). On the right the context image with umbral regions indicated by contours is displayed. The largest offsets are located on the edges of the umbral regions, where possible misalignment might occur during the data processing.}
\label{fig:offsets}
\end{figure*}

\begin{figure}
\plotone{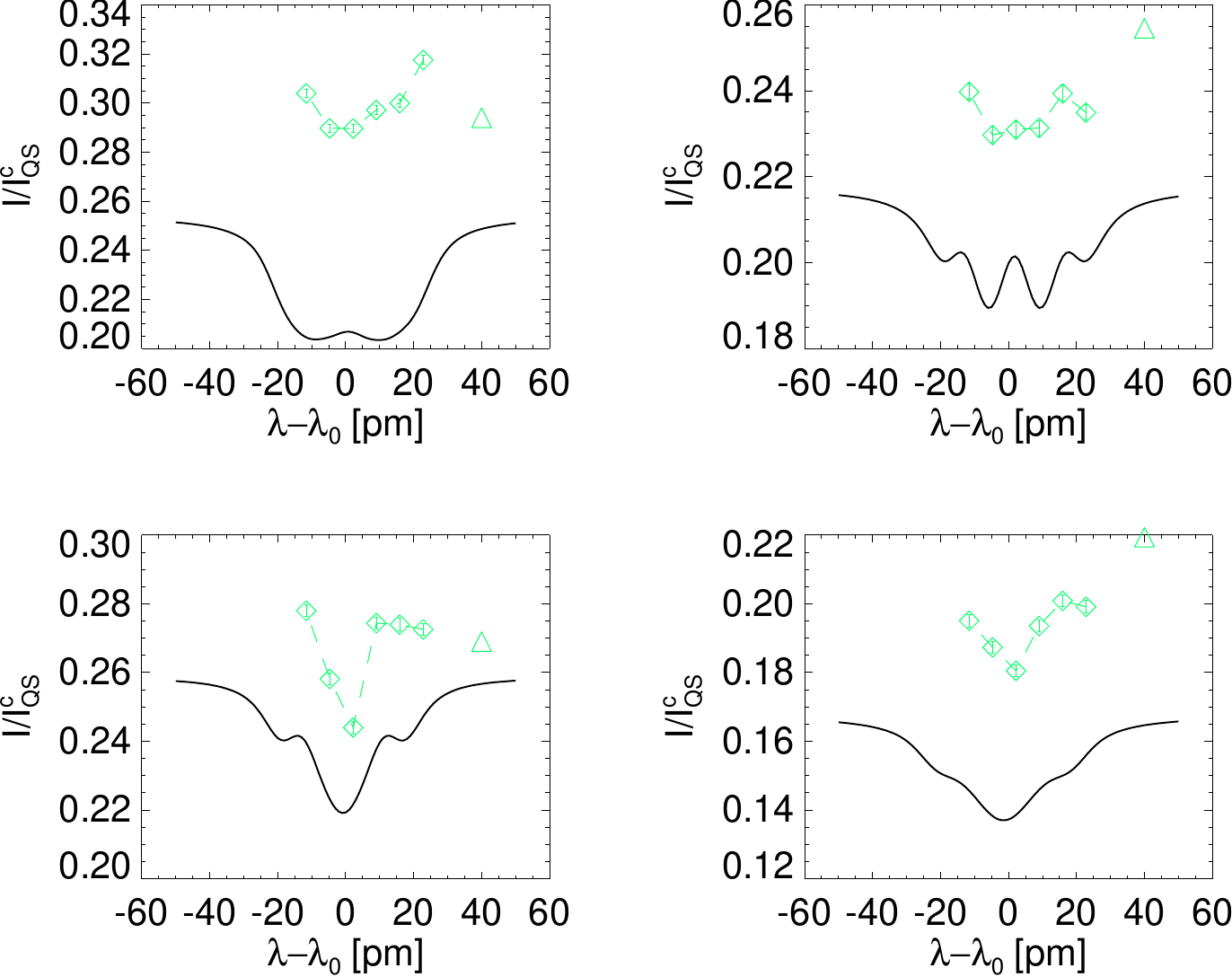}
\caption{Examples of umbral profiles where $\Ic$ has an obviously incorrect value either due to the broad or complicated line profile. The solid black line again represents the synthetised 617.3~nm line from atmospheres derived from inversions of Hinode/SP observations, green diamonds correspond to the spectral profile from HMI filtergrams, and the  green triangle denotes the HMI $\Ic$ pseudocontinuum. HMI has a positive offset with respect to the modelled line, however the offset for $\Ic$ is different in the displayed examples. }
\label{fig:profiles_umbraI}
\end{figure}

\begin{figure}
\plotone{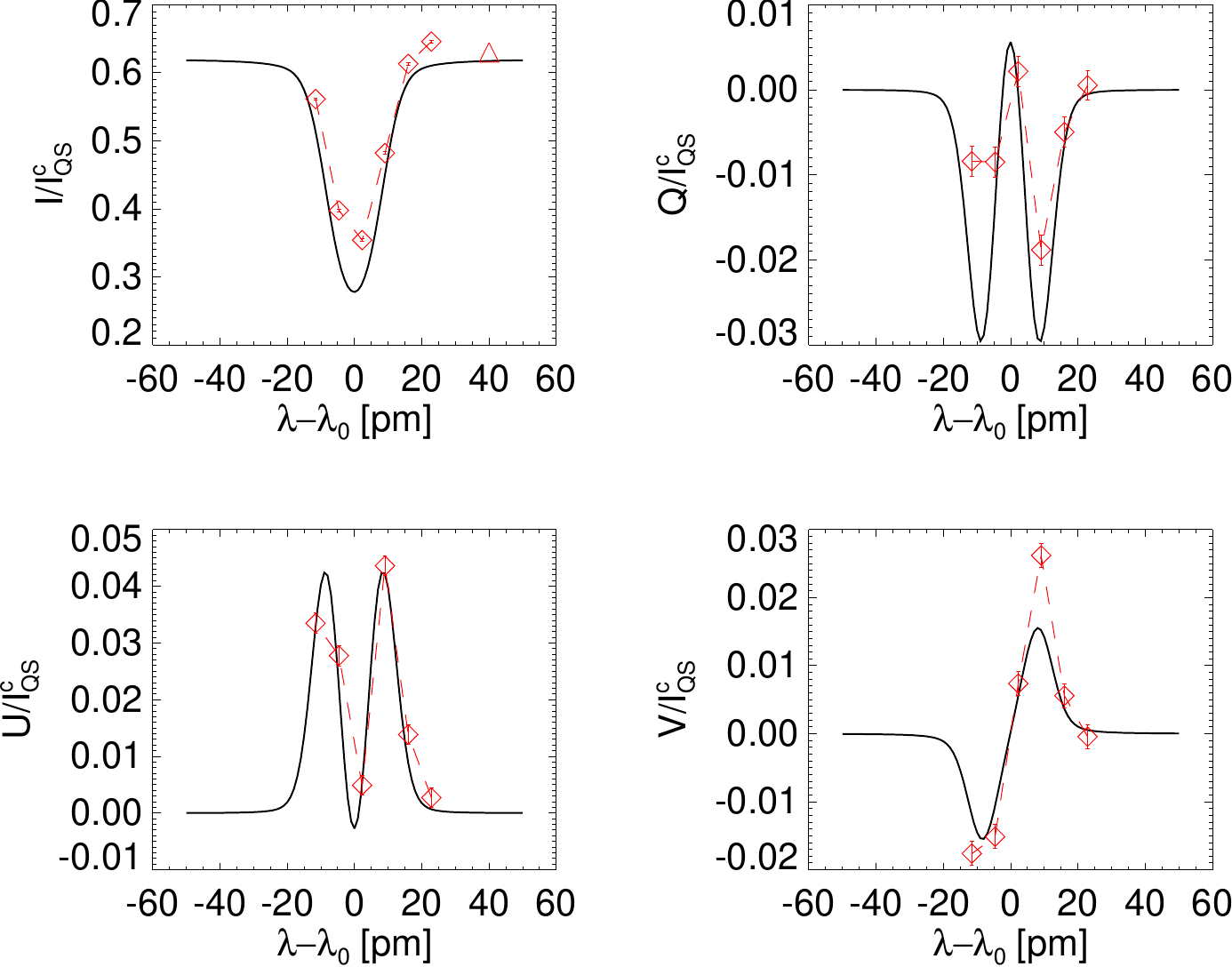}
\caption{The Stokes profiles for an example penumbral point. Symbols are the same as in Fig.~\ref{fig:profiles_quiet}.}
\label{fig:profiles_penumbra}
\end{figure}

\begin{figure}
\plotone{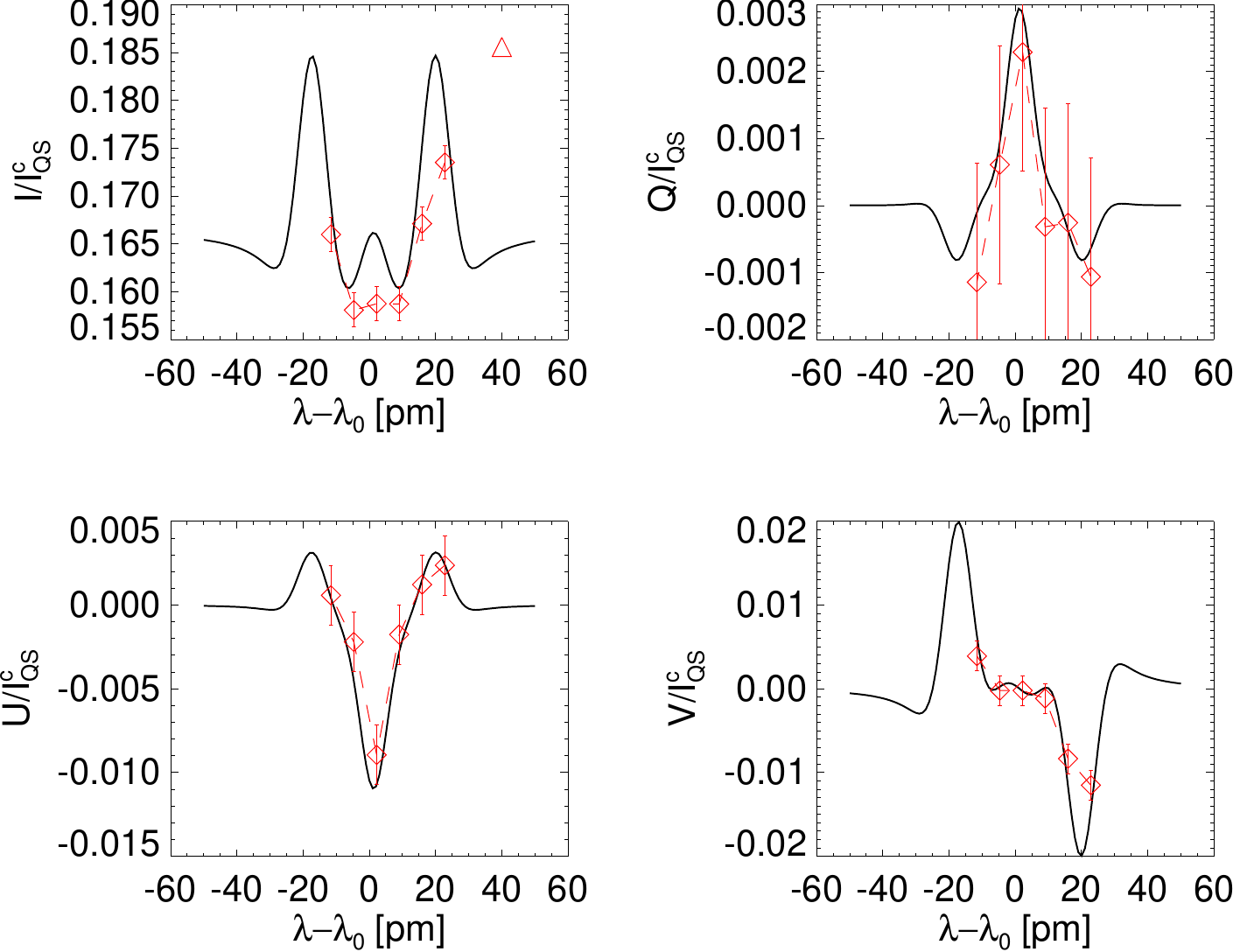}
\caption{The Stokes profiles for an example point showing line emission in $I$ profile. Symbols are the same as in Fig.~\ref{fig:profiles_quiet}.}
\label{fig:profiles_emission}
\end{figure}
\section{Results of comparison}\label{results}

Over the field-of-view we analysed all together 5475 pixels that were classified as reliable in terms of the $\chi^2$ value of the SIR inversion. For each of these pixels the synthetised $I$, $Q$, $U$, $V$ profiles are compared with the observed HMI profiles corrected by $A$ and $\Delta\lambda$ parameters as described in Sect.~\ref{data_processing}. Apart from that, %we did not perform any 
no other adjustment to the synthetic profiles was done. Some of the pixels show an excellent agreement between the two instruments, some do not. 

There is a variety of possible reasons for mismatch of the profiles from those two instruments. For instance, only a small spatial misalignment in a region where spatially confined features are present (such as granules vs. intergranular lanes or filaments in the penumbral regions) may lead to a significant change in the profile shapes. Similarly, a fast evolution may play a role (e.g. in pixels affected by a white-light flare ribbon during the accumulation time of HMI series). Therefore, we separately studied profiles in the quiet-Sun, penumbra, umbra, and the region where emission profiles are present, see Fig.~\ref{fig:FOV}. 

In the quiet-Sun regions, the synthetic profiles of the \ion{Fe}{i} 617.3~nm line and the HMI observations match quite well on average, as it is expected from the calibration method described in Sec.~\ref{data_processing}. Also, on average $\Ic$ represents the continuum level well. Any substantial differences are probably caused by misalignments both in space and time, where difference between the granules and intergranular lanes and their evolution plays a major role. Fig.~\ref{fig:profiles_quiet} shows an example of such a comparison. The synthetic Stokes profiles were smoothed by a Gaussian of FWHM=7.6~pm to approximate the real HMI transmission profiles. It seems that the six HMI spectral points correspond to a broader line than the smoothed modelled one, which is likely an effect of sidelobes in the real transmission profiles. The HMI spectral point outermost from the line centre is located already in the line continuum and thus may be used as a measurement of the nearby
continuum level (again, see Fig.~\ref{fig:profiles_quiet}). 

The umbral pixels were selected automatically by setting the continuum intensity threshold to $0.4~I^c_{QS}$ (continuum intensity of the surrounding quiet Sun). 
Furthermore, we excluded the pixels with line emission and the boundary pixels of continuous umbra regions to minimise the effects of misalignment on the umbra-penumbra or umbra-light bridge boundary. In total we analysed 381 umbral pixels. 

The main result of the umbral pixel analysis 
is that the intensities of HMI spectral points are systematically larger than predicted by the model, see Fig.~\ref{fig:profiles_umbra}. The histogram of the differences shows a Gaussian shape with a mean of about 4 per cent of the quiet-Sun intensity (see Fig.~\ref{fig:histograms}). A similar offset is observed also for the $\Ic$ values. 
Such an offset might easily be explained by scattered light in the instrument. The instrumental scattered light in HMI is not corrected by reduction routines \citep{2014SoPh..289.3483H}. On the other hand, there is a noticeable tail at larger offset values. From the offset maps (see Fig.~\ref{fig:offsets}) it is evident that the largest offsets are at the edges of umbral regions, where a spatial misalignment is possible due to the data processing, such as a leakage of the brighter intensity from the surrounding penumbral or light-bridge regions caused by interpolation during the registration of images. 

Additionally, the line profiles in the umbra are very wide in general. None of the HMI spectral points lie in the line continuum, they all sample the line core. Also, the spectral line profiles can have quite a complicated structure, i.e. far from a simple Gaussian shape. In some cases we see obvious $\sigma$-components 
of the Zeeman splitting. As a consequence, the MDI-like algorithm often does not determine a
continuum value that represents the real one. Depending on the real line profile in the pixel and the sampling points, the algorithm may both overestimate and underestimate the continuum value when compared with the modelled one, see Fig.~\ref{fig:profiles_umbraI}.

If the offset observed in the umbra is due to scattered light, then it should be present everywhere in the field of view. It cannot be determined for the quiet Sun, where it is hidden in the parameter $A$ (see Eq.~\ref{eq:fit}). However, it should be present in the penumbra. Indeed, the penumbral HMI $I$ profiles show 
a positive offset from the model profile. The histogram of penumbral intensity differences
is even broader than that related to umbral points, see Fig.~\ref{fig:histograms}. The reason is that the structures in the penumbra have much larger contrast, thus even a small misalignment leads to distinct profiles. The peak of the histogram is shifted towards positive values similarly to the histogram of the offsets in umbra. 

Regarding the $\Ic$ values, the line profiles in the penumbra are broader than in the quiet Sun, but narrower than in the umbra. 
Therefore, the HMI intensities at the outermost point from the line centre 
reach almost the continuum level in an extended wing of the line. Curiously, the $\Ic$ values are usually a bit below the expected continuum level, see Fig.~\ref{fig:profiles_penumbra}. This result is confirmed by the histogram of the offsets in the penumbra shown in Fig.~\ref{fig:histograms}. 
We assume that it is again an artifact of the simple MDI-like algorithm used to derive the continuum level. 

The last region to compare the line profiles is the region with emission profiles. Generally, the modelled profiles show a rather complicated structure, 
for an example see Fig.~\ref{fig:profiles_emission}, 
and if sampled only at six positions, their nature can be unresolved in HMI. Since most of the points in the emission are either in the umbra or in the light bridge, the line is broad here and thus even the HMI intensity at the outermost spectral point from the line core does not sample the continuum. Therefore, the $\Ic$ values are unreliable in the areas with line emission. We note a general enhancement of the modelled continuum, however its value may be overestimated by a large factor, if $\Ic$ is taken as its proxy. 
An example of such behaviour is given in Fig.~\ref{fig:profiles_emission}. 

In Figs.~\ref{fig:profiles_quiet}, \ref{fig:profiles_umbra}, \ref{fig:profiles_penumbra}, and \ref{fig:profiles_emission}, we display also the Stokes $Q$, $U$, and $V$ profiles observed by HMI and synthesised from model atmospheres based on Hinode SP observations. Typically, the agreement is not within the error bars for all spectral points but the overall shape is comparable. There are two effects that might contribute to the found discrepancies. First, the displayed error bars indicate 1$\sigma$ errors, the agreement would be better if one considers for instance 3$\sigma$ criterion. Second, the spectral line is not sampled simultaneously by HMI, therefore the time evolution of the fast-evolving features may play a role. The important conclusion is, however, that we do not observe an offset or systematic deviations in $Q$, $U$, and $V$ profiles. 

Using the direct comparison between the modelled continuum level with the $\Ic$ product indicates a large error in the determination of the latter. From histograms in Fig.~\ref{fig:histograms} one can see that the relative offsets in the umbra and penumbra are smaller for $\Ic$ than for the individual spectral points (the dashed curves are more to the left of the solid ones). It is the opposite for the emission region. The relative error  may be tens of per cent, see the values of relative offsets in Fig.~\ref{fig:histograms}. So, if one compares the $\Ic$ value in a given pixel before and during a flare, the contribution from the fit artefacts may be unacceptably large. 

It seems that $\Ic$ can be used to seek for localised temporal brightening \citep[the detection of white-light kernels, e.g.][]{2017NewA...57...14M}, but it is largely useless for an accurate retrieval of the light curve or any sort of photometric studies. Moreover, it is unclear to which extent the brightening in $\Ic$ indicates the enhancement in the continuum or whether it is an artefact of  inadequate sampling of the complicated line profile, i.e. an artefact of the line-core emission.

\section{Conclusions}
We directly compared the products based on observations by Hinode/SOT and HMI recorded during a very strong X-class flare. We use the opportunity to investigate the behaviour of Stokes spectral profiles in various regions of the field of view. Our aim was to learn about the $\Ic$ product available from HMI that is widely considered as an indicator for the photospheric continuum. We show that in the quiet-Sun regions the match between the Stokes profiles modelled from Hinode/SOT SP data and detected by HMI is acceptable and that the $\Ic$ value can serve as an indicator or a proxy for the continuum intensity. 

In the magnetised regions the situation is far less satisfactory. Even for pixels where a very good match between the Stokes profiles exists, the value of $\Ic$ may be strongly biased. This is due to the fact that all six HMI spectral sampling points are located within the line core broadened by the magnetic field. Therefore,  $\Ic$ is essentially extrapolated using a too simplistic model. 

The X9.3 flare of 6th September 2017 was also connected with an appearance of emission \ion{Fe}{i} profiles related to the flare ribbons. In these regions, it is  
possible to find pixels with a good match between modelled and HMI profiles, however in a large majority of pixels the real spectral-line shape is lost due to the inadequate sampling. The pseudocontinuum $\Ic$ is again strongly biased in these regions owing 
to the simplistic model for its calculation. 

We note that already \cite{2014SoPh..289.3483H} pointed out that the sparse and limited spectral sampling precludes a clean continuum point from being sampled consistently and that this issue together with the presence of scattered light can impact the interpretation of relative photometry especially for analysis of temporally varying features. The qualitative claim by \cite{2014SoPh..289.3483H} is thus quantitatively confirmed by our analysis. 

We stress that without the knowledge of the line profile (synthesised from an atmosphere resulting from inversions of Hinode data) it would not be possible to assess the $\Ic$ biases properly. The example in Fig.~\ref{fig:profiles_emission} shows a case when a profile with a strong emission is sampled by the six HMI points in such a way that one would consider this line to be a simple absorption line. Yet, the model reveals clear emission in the $I$ and $Q$, $U$, $V$ profiles consistent with the observed ones.

We conclude that HMI's $\Ic$ may be used as an indicator for the propagation of flare ribbons. However, without knowing the real profile of the line one cannot simply say whether the derived enhancement is an enhancement of the continuum or whether it is an artifact of the complex shape of the HMI spectral line. Thus, $\Ic$ must be used with caution in any kind of photometric or energetic studies of flares.

\acknowledgments
We thank the HMI team, in particular Yang Liu, for their efforts to calibrate and to provide new data products of HMI. 
The authors were supported by Czech Science Foundation under grants 18-06319S (M\v{S} and JJ) and 16-18495S (JK). The support by project RVO:67985815 is also acknowledged. Hinode is a Japanese mission developed and launched by ISAS/JAXA, with NAOJ as domestic partner and NASA and STFC (UK) as international partners. It is operated by these agencies in cooperation with ESA and NSC (Norway).  %The HMI data are courtesy of NASA/SDO and the HMI science team.

% \bibliographystyle{aasjournal}
% \bibliography{biblio}

\end{document}